 \documentstyle[prl,aps,multicol,epsfig]{revtex}
\begin{document}
\draft \date{\today} \title
  {Why an ac magnetic field shifts the irreversibility line
   in type-II superconductors}

\author{Ernst Helmut Brandt$^1$ and Grigorii P.~Mikitik$^{2,1}$}

\address{$^1$Max-Planck-Institut f\"ur Metallforschung,
   D-70506 Stuttgart, Germany\\
         $^2$Verkin Institute for Low Temperature Physics \&
   Engineering, Ukrainian Academy of Sciences,
   Kharkov 61103, Ukraine}

\maketitle

\begin{abstract}
We show that for a thin superconducting strip placed in a
transverse dc magnetic field - the typical geometry of experiments
with high-$T_c$ superconductors - the application of
a weak ac magnetic field perpendicular to the dc field generates a
dc voltage in the strip. This voltage leads to the decay of the
critical currents circulating in the strip, and eventually the
equilibrium state of the superconductor is established. This
relaxation is not due to thermally activated flux creep but to the
 ``walking'' motion of vortices in the {\it two-dimensional}
  critical state of the strip with in-plane ac field.
Our theory explains the shaking effect that was used for detecting
phase transitions of the vortex lattice in superconductors with
pinning. Some recent experiments on this subject are discussed.
\end{abstract}
    \begin{multicols}{2}
    \narrowtext

  Experimental investigation of the equilibrium properties of
type-II superconductors can be performed only in the reversible
region of the magnetic field ($H$) versus temperature ($T$) plane.
But often strong flux-line pinning prevents the measurement of
equilibrium properties. In this context, Willemin et al.\cite{1}
recently made an interesting observation. Their experiments
revealed that application of an additional small oscillating
magnetic field {\it perpendicular} to the main dc field leads to a
fast decay of the currents circulating in the critical state of
various high-$T_c$ superconductors. This effect dramatically
extends the observable reversible domain in the $H$-$T$ plane. The
relaxation of the irreversible magnetization in these experiments
was exponential in time, and thus was obviously different from
thermally activated flux creep, which leads to a logarithmic time
law. Using this vortex-shaking process, the melting transition of
the vortex lattice in ${\rm YBa_2Cu_3O}_{7-\delta}$ crystals was
detected\cite{2} at temperatures very close to the critical
temperature $T_c$, where the melting could not be investigated
before.  With the same shaking method, it was discovered \cite{3}
that the order-disorder transition in the vortex lattice of ${\rm
Bi_2Sr_2CaCu_2O_8}$ is of the first order at low temperatures,
where vortex pinning usually masks the corresponding jump in the
equilibrium magnetization. Thus, the vortex-shaking process
opens new possibilities in experimental investigation of the
$H$-$T$ phase diagrams of superconductors. Nevertheless, the
nature of this important effect so far remained unclear.

   In this paper we give a quantitative explanation of this
vortex-shaking effect for thin strips. It is shown that the
relaxation is caused by the generation of a {\it dc electric
field} by the {\it ac magnetic field}. We obtain this result in
the framework of a quasistatic approach based on the standard
critical state concept. Therefore, this effect is quite general
and should occur in all type-II superconductors, not only in
high-$T_c$ materials. In contrast to the generation of an electric
field in the case\cite{4} when the ac and the dc magnetic fields
are {\it parallel}, the above electric field appears in
superconductors {\it without} a transport current. We also point
out a new way of measuring the critical current density
$j_c$ in superconductors.

   The dc electric field in general depends on the orientation
of the ac magnetic field with respect to the currents
circulating in the superconductor. In the following we shall
analyze the simplest situation: A thin superconducting strip fills
the space $|x| \le w$, $|y| < \infty$, $|z| \le d/2$ with $d\ll
w$; the constant and homogeneous external magnetic field $H_a$ is
directed along $z$, while the ac magnetic field $h \cos \omega t$
is applied along $x$, i.e., perpendicular to $H_a$ and to the
currents in the sample (Fig.~1). We also make here the
usual Bean assumption that the critical current density $j_c$ does
not depend on the local induction $B$. The field $H_a$ is assumed
to be sufficiently large to exceed both the field of full
penetration for the strip,\cite{5} $H_p = (j_c d /\pi) \ln (2 e
w/d)$, and the lower critical field $H_{c1}$, and so we may put $B
= \mu_0 H$.

A qualitative explanation of the shaking effect is as follows:
The currents flowing in the critical state of the strip generate
a nonuniform distribution of the magnetic induction $B_z(x)$.
The ac field periodically tilts the vortices in this state.
However, at each point $x$ with a nonzero sheet current $J(x)$
(the current density integrated over the thickness $d$), the
tilt is {\it not symmetric} relative to the central plane of the
strip $z=0$, and during each cycle of the ac field, the asymmetry
leads to a shift of vortices towards the center $x=0$ of the strip.
This shift tends to equilibrate $B_z(x)$, and it also generates
a dc electric field which decreases $J(x)$. When $J(x)$ reaches
zero, the asymmetry disappears, and the process stops.

It will be shown below that the generated dc electric field is
proportional to the thickness of the sample, $d$. Thus, to
describe the effect, the strip cannot be considered as infinitely
thin, and the appropriate critical state problem is two
dimensional. However, according to the approach of
Ref.~\onlinecite{6}, the smallness of the parameter $d/w$ enables
us to {\it split the problem into two simpler problems\/}: A
one-dimensional problem across the thickness of the sample, and a
problem for the infinitely thin strip. Namely, we first interpret
a small section of the strip around an arbitrary point $x$
(see Fig.~1) as an ``infinite''
slab of thickness $d$ placed in a perpendicular dc magnetic field
$B_z(x)$ and in a parallel ac field $h \cos \omega t$ and
carrying a sheet current $J(x)$. The resulting  dc electric field
$E_y=E_y(J,B_z,h)$ for the slab we then use as the local electric
field $E_y(x)$ for an infinitely thin strip, to calculate the
temporal evolution of the sheet current $J(x)$ and induction
$B_z(x)$ in this strip by the method of Ref.~\onlinecite{7}.

   We begin with the analysis of the dc electric field $E_y$
generated by the ac magnetic field, $h\cos\omega t$, in an
infinite slab with sheet current $J$ and with a constant and
homogeneous magnetic field $B_z$. At sufficiently small
$\omega$ \cite{8}, the problem may be considered quasistatically:
We solve the critical state equation for the slab,
$dB_x/dz = \mu_0 j_{cr}(z,t)$, at every moment of time $t$, and
then find the shape of the flux lines at this moment from
$dx/B_x=dz/B_z$ and the condition that the points on each line
where $j_{cr}$ changes its sign cannot move. The critical current
density $j_{cr}$ is always equal to $j_c$ or $-j_c$, and its
distribution over $z$ is just the well-known distribution in a
slab in a {\it parallel} magnetic field and with current $J$.
In particular, at $t=0$ and $t=2\pi/\omega$
one has $j_{cr}=j_c$ for $z_{-}<z\le d/2$, and $j_{cr}=-j_c$
for $-d/2\le z<z_{-}$ (see Fig.~1), while at $t=\pi/\omega$ one
has $j_{cr} =-j_c$ for $z_{+}<z\le d/2$, and $j_{cr}=j_c$ for
$-d/2\le z <z_{+}$ where $z_+=-z_-=J/2j_c$. An essential finding
of this analysis is that during one half cycle
($0\le t\le \pi/\omega$) every flux line turns
around a fixed ``swivel point'' with the coordinate $z=z_+$, while
during the second half cycle ($\pi/\omega\le t\le 2\pi/\omega$)
the line turns around another fixed point with $z=z_-$. As a
result, the line ``walks'' along $x$, see Fig.~1. The shift of the
vortices during one full cycle is
\begin{equation} 
\Delta x={2\mu_0 J \over j_c B_z}[\, h-h_p(J) \,] ,
\end{equation}
where $h_p(J)=(J_c-|J|)/2$ is the field of
 full penetration of {\it parallel} flux
into a slab with current $J$, and $J_c=j_c d$ is the critical
sheet current. Note that in deriving $\Delta x$ we have assumed
$h\ge h_p(J)$. Else, the $x$ component of the magnetic field
does not penetrate completely into the slab, a section of the flux
line does not sway and hence does not move at all,
and thus $\Delta x = 0$. The dc
electric field generated by the above shift of the vortex lines
is $E_y=(\omega/2\pi)\Delta x B_z$, and hence we arrive at
  \begin{eqnarray}  
  E_y =& 0  ~~~~~~~~~~~~~~~~~~~
                        &{\rm for}~~ h < h_p(J) \,, \nonumber \\
  E_y =& {\displaystyle
             {\mu_0 \omega d J \over \pi J_c} } [h - h_p(J)]
                        ~~~~&{\rm for}~~ h \ge h_p(J) \,.
  \end{eqnarray}

The described mechanism for the generation of a dc electric field
was proposed many years ago\cite{8}, and Eq.~(2) is equivalent to
the formula (2.8) of Ref.~\onlinecite{8}. Equation~(2) also
coincides with the corresponding result for the slab obtained in
our paper\cite{4}, though in Ref.~\onlinecite{4} the dc magnetic
field lies in the plane of the slab. This coincidence\cite{9} is
not surprising, since in both cases $E_y$ may be interpreted as
generated by the transfer of the $x$ component of the flux across
the slab thickness $d$.

In the derivation of $E_y$ we have assumed that $J$ and $B_z$ do
not change during one cycle. This approximation is justified when
the relaxation time of the profiles $J(x)$ and $B_z(x)$
considerably exceeds the period $2\pi/\omega$ of the ac field. As
will be shown below, this condition is indeed fulfilled for thin
samples, $d\ll w$, and thus the approximation is self-consistent.

We now consider the temporal evolution of the profiles $J(x)$ and
$B_z(x)$ in the infinitely thin strip.\cite{7} This evolution is
caused by the slow ``walking'' of vortices toward the center of
the strip and is given by the Maxwell equation:
 \begin{equation} 
  {\partial B_z \over \partial t}=-{\partial E_y \over \partial
  x},
 \end{equation}
with $E_y$ specified by formula (2). Since $E_y=v_x B_z$, where
$v_x$ is the average velocity of the vortices, the above equation
simply expresses the conservation law of their number inside the
strip. On the other hand, the Biot-Savart law yields a relation
between the sheet current $J$ and $B_z$:
   \begin{equation} 
   {B_z(x) \over \mu_0}=H_a + {1 \over 2\pi} \int_{-w}^w {J(u)du \over
   u-x}\ .
  \end{equation}
Equations (2)-(4) together with the initial condition
  \begin{equation} 
  J(x)|_{t=0}=-J_c x / |x| \,,
  \end{equation}
are sufficient to determine the temporal evolution of $J$, $E_y$,
and $B_z$ in the strip, i.e., the functions $J(x,t)$, $E_y(x,t)$,
and $B_z(x,t)$.

To proceed further, it is convenient to rearrange Eq.~(3), (4) as
follows: We invert\cite{10} Eq.~(4), differentiate both sides
 of the result
with respect to $t$, insert Eq.~(3), and arrive at an equation for
$J(x,t)$:
 \begin{equation} 
  {\partial J(x,t) \over \partial t}={2 \over \pi \mu_0}
  \int_{-w}^w \!
  {du \over  u\!-\!x} \left({w^2\!-\!u^2 \over w^2\!-\!x^2}\right)^{\!\!1/2}
  \! {\partial E_y(J) \over \partial u}
 \end{equation}
with $E_y(J)$ given by formula (2). The
right hand side of Eq.~(6) is proportional to $(d/w)\omega$; this
becomes evident if one introduces the dimensionless length $x/w$.
Thus, the decrease of $J/J_c$ during one cycle is determined by
the small parameter $d/w$, and the above assumption that $J$ and
$B_z$ are constant over one cycle is well justified.

The numerical method of solving Eq.~(6) is well
elaborated,\cite{7} and we use it to analyze the evolution of the
sheet current $J(x,t)$. In Fig.~2 we show the time dependence of
the magnetic moment \cite{11} per unit length of the strip,
$M(t) = \int_{-w}^w J(x,t) x\, dx$, at various amplitudes
$h$ of the the ac magnetic field. If $ h/J_c  > 0.5$, then
$|M(t)|$ decreases almost exponentially with time, and
eventually
 the equilibrium state\cite{12}
is established since $J \to 0$
everywhere in the strip; when $h/J_c < 0.5$, the equilibrium state
is not reached, Fig.~3. At large times $t$, the
saturation value of the current in the sample, $J_{\infty}$, can
be found from the condition  $E_y(J_{\infty}) = 0$, yielding
\begin{equation} 
  J_{\infty}=J_c-2h\ .
\end{equation}
In the special case $h =h^* =0.5J_c$, Eq.~(2) gives $E_y \propto
J^2$, and it follows from Eq.~(6) that $M\propto J \propto
t^{-1}$. Thus, we conclude the following: When in an experiment
the amplitude $h$ is fixed while $j_c$ increases due to a decrease
of temperature $T$, then at some $T$ one has $J_c=2h$, and at
lower $T$ the vortex-shaking will not lead to the equilibrium
state. In other words, the greater is $j_c d$, the greater is the
ac amplitude $h$ required for the equilibrium state to be reached.
This feature of the vortex-shaking process was indeed observed in
the experiments.\cite{2,3}

  We consider now more closely the practically important
case $h>J_c/2$. When $J$ decays, then, according to Eq.~(2),
the problem formally
reduces to the Ohmic strip problem analyzed in
Ref.~\onlinecite{7}, but with the resistivity now associated
neither with free flux flow nor with thermally assisted flux flow,
 namely, $E/j =\rho =\mu_0 \nu d^2 (2h/J_c-1)=$ const,
$\nu=\omega/2\pi$.
In this case the sheet current after some transient time tends to
the following solution of Eq.~(6):
\begin{equation}  
  J(x,t) =-C J_c\, f(x/w) \exp(-t/\tau) \,,
\end{equation}
where $C$ is a constant depending on the magnetic history,
$f(v)$ is the normalized eigenfunction
with lowest eigenvalue $\Lambda$ of the integral equation,
 \begin{equation} 
  \Lambda f(v) =-{1 \over \pi^2}  \int_{-1}^1 \!
  {du \over  u-v} \left({1-u^2 \over 1-v^2}\right)^{\!\!1/2}\!
  {d f(u) \over d u} \,,
 \end{equation}
and the relaxation time $\tau$ is given by
 \begin{equation} 
  \tau^{-1} =  \Lambda {\omega d \over  w }
  \left({2h-J_c \over J_c}\right)\,.
 \end{equation}
Numerical analysis\cite{7} yields $\Lambda = 0.6386$.
The odd function $f(v)=-f(-v)$ normalized to
$\int_0^1\! v f(v)dv ={1\over2}$ is shown in Fig.~4.
The magnetic moment per unit length of the strip,
$M= \int_{-w}^w\! x J dx$, is then $M=-C w^2 J_c \exp(-t/\tau)$,
while in the fully penetrated critical state one has
$M = M_0 = -w^2 J_c$ and in the Meissner state\cite{7}
$M = -\pi w^2 H_a$.

Thus, we obtain that at $h > J_c/2$ the decay of the current
becomes exponential in time, in agreement with the experimental
data\cite{1}, and the spatial profile of the current tends to the
universal function $f(x/w)$, which is independent of the
superconducting parameters and dimensions of the strip. It also
follows from formula (10) that the rate of the decay increases
with $\omega$ and $h$. However, for the above quasistatic analysis
to be valid, $\omega$ should not be too large\cite{8}, while the
condition $(\tau \omega/2\pi)\gg 1$ leads to $(w/d)(J_c/h)\gg 1$,
or $h\ll j_c w$. This puts an upper bound on $h$ up to which the
above formulas are applicable.

   Interestingly, with Eqs.~(7) and (10), measurements of the
amplitude $h^*$ at which $J_{\infty}=0$, or of $\tau$ for $h>h^*$,
allow one to obtain $j_c$ at short times $t \sim \omega^{-1}$
where $j_c$ is only weakly affected by thermally activated creep.

To conclude, we give a quantitative description of the
vortex-shaking effect in a thin superconducting strip. It is shown
that in the framework of the critical-state theory, the decay of
the critical currents to zero, or the decrease of the magnetic
moment of the strip to its equilibrium value,
is due to the generation of a
dc electric field by an ac magnetic field applied normal to the
main dc magnetic field. This electric field originates only if the
finite thickness of the strip is taken into account, and so the
theoretical description of the vortex-shaking effect is, strictly
speaking, a two-dimensional spatial problem. But the small value
of the parameter $d/w$ enabled us to simplify the problem and to
solve it explicitly for ac amplitude $h\ll j_c w$.
When the thin superconductor has a different shape, e.g.\ looks
like a square platelet, there are regions where the ac magnetic
field is parallel to the sheet current. In this case, the analysis
of the problem becomes more complicated. Nevertheless,
the obtained results explain the properties of the vortex-shaking
effect observed in the experiments\cite{1,2,3} and also shed light
on the origin of similar intriguing effects\cite{13,14,15}
discovered earlier for slightly different geometries.

\vspace{-0.3 cm} 

 \begin{figure}[F1]
\epsfxsize= .95\hsize  \vskip 1.0\baselineskip \centerline{
\epsffile{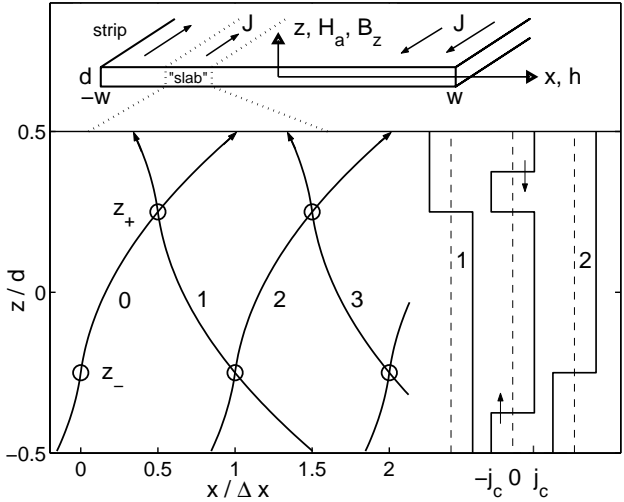}}
  \vspace{.2cm}
\caption{Strip geometry and magnetic fields (upper inset).
A flux line ``walking'' from left to right through a section of
the strip (``slab'') is shown at times $t\omega/\pi = 0$, 1, 2, 3.
The swivel points of the line are marked by
circles. Here $h/J_c =0.6$, $J/J_c =0.5$, yielding $z_+ =-z_- =
0.25 d$ and $\Delta x = 0.35 (\mu_0 J /B_z) d$. $\Delta x$ is the
shift per cycle, Eq.~(1); $x$ is measured from an arbitrary point
of the ``slab''. The line shape consists of parabolas with
$d^2x/dz^2 = \pm \mu_0 j_c / B_z$. The scheme at the right shows
the current profiles across the slab at the extremal times 1 and 2
and at some intermediate time, where the two arrows indicate the
penetration of the current-inversion front.
 } \end{figure}
\vspace{-.5cm}

 \begin{figure}[F2]
\epsfxsize=.94\hsize  \vskip 1.0\baselineskip \centerline{
\epsffile{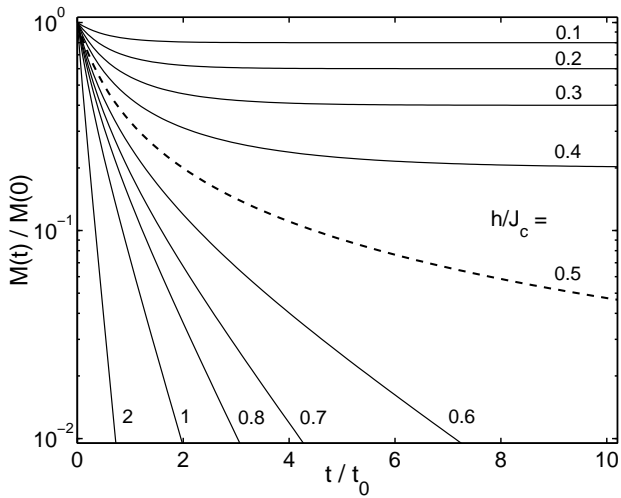}}
 \vspace{.2cm}
\caption{The time dependence of the magnetic moment $M(t)$ of the
strip at various amplitudes $h$ of the ac magnetic field: $h/J_c$
= 0.1, 0.2, $\dots$ 0.8, 1, 2. The dashed line ($h/J_c=0.5$)
separates the complete and incomplete relaxation of $M$. $M(0) =
M_0 = -w^2 J_c$ is the magnetic moment of the strip in the fully
penetrated critical state. The time unit $t_0 = (\pi w / \omega
d)$ corresponds to $w/2d \gg 1$ cycles.
 } \end{figure}

 \begin{figure}[F3]
\epsfxsize= .85\hsize  \vskip 1.0\baselineskip \centerline{
\epsffile{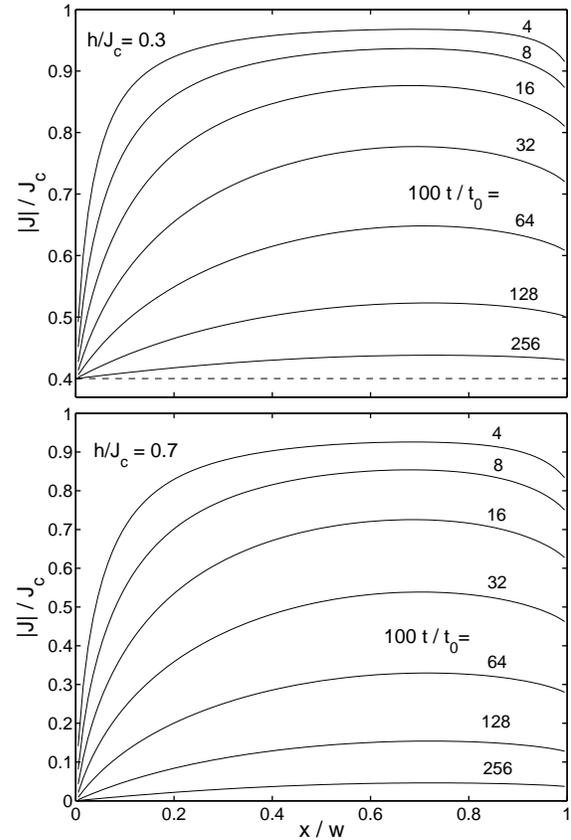}}
 \vspace{.1cm}
\caption{Profiles of the sheet current $J(x,t)$ in the strip at
various times $t$ (in units $t_0/100$, $t_0 =\pi w/\omega d$)
starting from Eq.~(5). Top: $h/J_c = 0.3$; the dashed line shows
$J_\infty$, Eq.~(7). Bottom: $h/J_c=0.7$; exponential relaxation
to $J=0$.
 } \end{figure}

\begin{figure}[F4]
\epsfclipon \epsfxsize= .90\hsize  \vskip 1.0\baselineskip
\centerline{\epsfclipon \epsffile{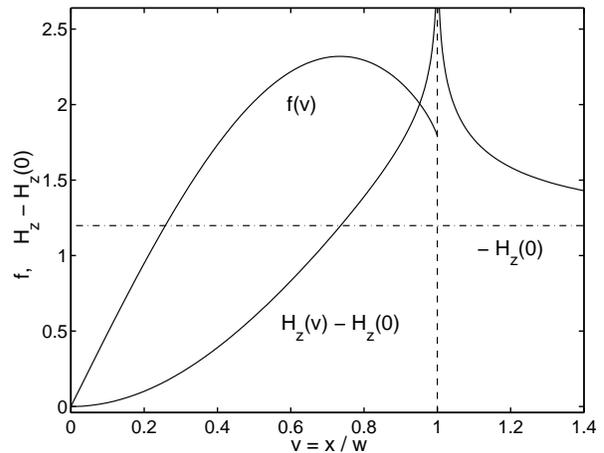}}
 \vspace{.1cm}
\caption{The universal spatial profile $f(v)$ of the relaxing
sheet current $J(x,t)$ in the strip at $t > \tau$, Eq.~(8), and
the magnetic field $H_z$ generated by this current profile. The
dash-dotted line indicates $-H_z(0)$. $H_z$ is in units $J/f = C
J_c \exp(-t/\tau) = -M(t)/w^2$. At the edges $H_z$ has a
logarithmic infinity, which is cut off by the finite thickness
$d$.
 } \end{figure}

\end{multicols}
\end{document}